\documentclass{cimento}

\usepackage{graphicx}
\usepackage{amssymb}

\title{Assessing the Predictive Power of Galaxy Formation Models with 
the Rest-Frame Optical Luminosity Functions at $2.0 \leq z \leq 3.3$}

\author{Danilo~Marchesini\from{ins:yale} \atque
Pieter G. van Dokkum\from{ins:yale}}

\instlist{\inst{ins:yale} Department of Astronomy - Yale Center for Astronomy and Astrophysics, Yale University, New Haven, CT, USA}

\PACSes{\PACSit{98.62.Ve}{Characteristics and properties of external galaxies and extragalactic objects -- statistical and correlative studies of properties}}

\shorttitle{LFs at $z \geq 2$: observations vs. predictions}
\shortauthor{Marchesini \& van Dokkum}

\begin{document}

\maketitle

\begin{abstract}
We compare recently measured rest-frame $V$-band luminosity 
functions (LFs) of galaxies 
at redshifts $2.0 \leq z \leq 3.3$ to predictions of semianalytic 
models by De Lucia \& Blaizot and Bower et al. and hydrodynamic 
simulations by Dav\'e et al. The models succeed for some 
luminosity and redshift ranges and fail for others. A notable 
success is that the Bower et al.\ model provides a good match to 
the observed LF at $z\sim 3$. However, all models predict an 
increase with time of the rest-frame $V$-band luminosity density, 
whereas the observations show a decrease. The models also have 
difficulty matching the observed rest-frame colors of galaxies. 
In all models the luminosity density of red galaxies increases 
sharply from $z\sim3$ to $z\sim2.2$, whereas it is approximately 
constant in the observations. Conversely, in the models the 
luminosity density of blue galaxies is approximately constant, 
whereas it decreases in the observations. These discrepancies cannot 
be entirely remedied by changing the treatment of dust and suggest 
that current models do not yet provide an adequate description of 
galaxy formation and evolution since $z\sim 3$.
\end{abstract}


\section{Introduction}

In the current paradigm of structure formation, dark matter (DM) halos
build up in a hierarchical fashion through the dissipationless
mechanism of gravitational instability. The assembly of the stellar
content of galaxies is instead governed by much more complicated
physical processes, often dissipative and non-linear, which are
generally poorly understood. To counter this lack of understanding,
prescriptions are employed in the galaxy formation models. One of the
fundamental tools for constraining the physical processes encoded in 
these models is the luminosity function (LF), since its shape retains 
the imprint of galaxy formation and evolution processes.

The faint end of the LF can be matched with a combination of supernova
feedback and the suppression of gas cooling in low-mass halos due to a
background of photoionizing radiation (e.g. \cite{benson02}). Matching 
the bright end of the LF has proven more challenging. Very recent 
implementation of active galactic nucleus (AGN) feedback in semianalytic 
models (SAMs) has yielded exceptionally faithful reproductions of the 
observed local rest-frame $B$- and $K$-band global LFs 
\cite{bower06,croton06}, including good matches to the local rest-frame 
$B$-band LFs of red and blue galaxies (although with some discrepancies 
for faint red galaxies; \cite{croton06}).

The excellent agreement between observations and models at $z\sim0$ is
impressive but is partly due to the fact that the model parameters were
adjusted to obtain the best match to the local universe. A key
question is therefore how well these models predict the LF at earlier
times. The SAMs of \cite{bower06}, \cite{croton06}, and \cite{delucia06} 
have been compared to observations at $0<z<2$ \cite{bower06,kitzbichler06}. 
Although the agreement is generally good, \cite{kitzbichler06} infer that the
abundance of galaxies near the knee of the LF at high redshift is
larger in the SAMs than in the observations (except possibly for the
brightest objects), in an apparent reversal of previous studies (e.g.
\cite{cimatti02}). 

We have compared the observed rest-frame optical LFs of galaxies at 
$2.0 \leq z \leq 3.3$ \cite{marchesini07} to those predicted by 
theoretical models in this redshift range, in order to test the 
predictive power of the latest generation of galaxy formation models. 
We have also compared the observed LF to predictions from smoothed 
particle hydrodynamics (SPH) simulations, which have so far only been 
compared to data at $z\sim6$ \cite{dave06}. The details of this 
comparison can be found in \cite{marchesini07b}; here we summarize 
the main results from our study. All magnitudes are in the AB system, 
while colors are on the Vega system.


\section{The observed luminosity functions} \label{olf}

The observed rest-frame optical LFs at $z \geq2$ were measured 
from a composite sample of galaxies built from three deep 
multi-wavelength surveys, all having high-quality optical to 
NIR photometry: the ultradeep
 Faint InfraRed Extragalactic Survey (FIRES; \cite{franx03}), 
the Great Observatories Origins Deep Survey (GOODS; 
\cite{giavalisco04}; Chandra Deep Field--South), and the 
MUltiwavelength Survey by Yale-Chile (MUSYC; \cite{quadri06}). 
The unique combination of surveyed area and depth of the MUSYC 
survey allowed us to (1) minimize the effects of sample 
variance, and (2) better probe the bright end with unprecedented 
statistics. The FIRES allowed us to constrain the faint 
end of the LF, while the CDFS catalog bridges the two slightly 
overlapping regimes and improves the number statistics. The 
final $K$-selected sample, comprising a total of $\sim$990 
galaxies\footnote{Of these, $\sim$4\% have spectroscopic redshifts.} 
with $K_{\rm s}^{\rm tot}<25$ at $2 \leq z \leq 3.5$ over an area 
of $\sim$380~arcmin$^{2}$, is described in details in 
\cite{marchesini07}. The rest-frame optical ($B$, $V$, and $R$ bands) 
LFs of galaxies at $2.0<z \leq 3.5$ and a comprehensive analysis of 
the systematics effects due to photometric redshift uncertainties are 
also presented in \cite{marchesini07}. By splitting the sample by color, 
we showed that the LFs of red and blue galaxies are significantly 
different, with the latter being characterized by a much steeper 
faint-end slope. We also estimated the contribution of red and blue 
galaxies to the global densities. While blue galaxies dominate the 
rest-frame optical luminosity density, the global stellar mass density 
at $2< z< 3.5$ appears to be dominated by red galaxies.

Here, we limit our comparison between observed and predicted LFs 
to the rest-frame $V$ band, at the two redshift intervals 
$2.7\leq z \leq 3.3$ and $2\leq z \leq 2.5$.


\section{The model-predicted luminosity functions} \label{plf}

The SAM of Bower et al. \cite{bower06} is implemented on the 
Millennium DM simulation described in \cite{springel05}. The 
details of the assumed prescriptions and the specific parameter 
choices are described in \cite{bower06,cole00,benson03}. We have 
also used the outputs from the SAM of Croton et al. \cite{croton06} 
as updated by De~Lucia \& Blaizot \cite{delucia06}. This model 
differs from the SAM of Bower in the scheme for building the merger 
trees and in the prescriptions adopted to model the baryonic physics, 
most notably those associated with the growth of and the feedback 
from SMBHs in galaxy nuclei and the cooling model (see 
\cite{delucia06,kauffmann00,springel01,delucia04} for details). 
Finally, we have compared the observed LFs with the predictions 
from the cosmological SPH simulations of \cite{oppenheimer06}, 
already used in \cite{finlator06} to constrain the physical 
properties of $z \sim 6$ galaxies. The key ingredient of these 
simulations is the inclusion of superwind feedback. Specifically, 
we used the ``momentum-driven wind'' model used in \cite{finlator06} 
(see \cite{oppenheimer06} for detailed descriptions). A description 
of how the SAM-predicted rest-frame $V$-band LFs were computed 
can be found in \cite{marchesini07b}.


\section{Results} \label{results}

\begin{figure}
\centering
\includegraphics[height=11cm,width=7cm]{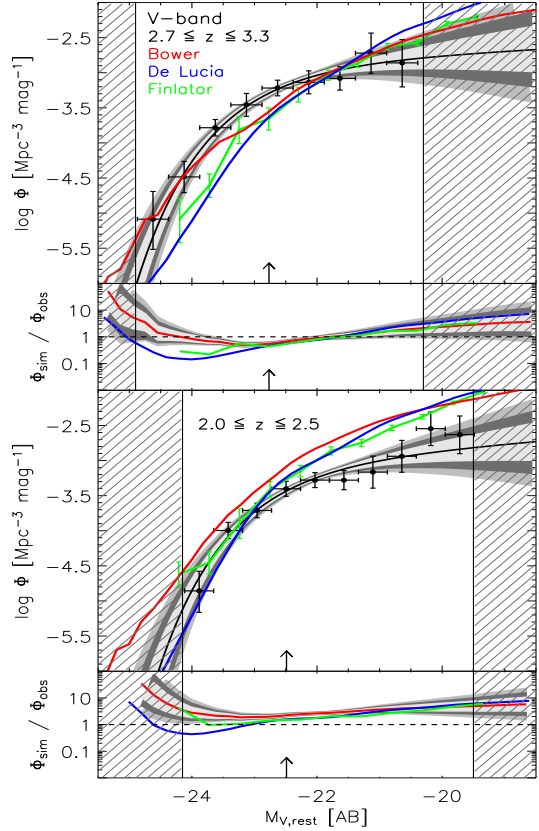}
\caption{From \cite{marchesini07b}: Comparison between the 
rest-frame $V$-band observed global LFs and those predicted by 
models. The observed LFs are plotted with black circles 
($1/V_{\rm max}$ method; \cite{avni80}) with 1~$\sigma$ error 
bars (including field-to-field variance) and by the black solid 
line (maximum likelihood method; \cite{sandage79}) with 1, 2, 
and 3~$\sigma$ solutions ({\it gray shaded regions}). The arrow 
shows the observed value of $M^{\star}$. Red lines show predictions 
from the SAM of \cite{bower06}, blue lines from the SAM of 
\cite{delucia06}, and green lines from the SPH model of 
\cite{finlator06}. Poisson errors ($1~\sigma$) are shown for the 
SPH model only, as they are very small for the SAMs. In the small
panels, the ratio between the predicted and the observed LFs is
plotted, together with the 1, 2, and 3~$\sigma$ errors for the SAM of 
\cite{bower06} ({\it gray shaded regions}). The oblique line regions 
delimit the comparison to the luminosity range probed by the sample 
of \cite{marchesini07}.
\label{fig1}}
\end{figure}

The comparison between the observed rest-frame $V$-band LFs of 
all galaxies at $2.7 \leq z \leq 3.3$ and $2 \leq z \leq 2.5$ 
with those predicted by the theoretical models is shown in 
Figure~\ref{fig1}. It is immediately obvious that the models do 
not yet provide a precise description of galaxy evolution. 
Differences between the various models, and discrepancies between 
model predictions and data, are still as large as a factor of 
$\sim 5$ for certain luminosity and redshift ranges.

At $2.7 \leq z \leq 3.3$, the global LF predicted by the SAM of
Bower agrees well with the observed LF. However, while at 
$2 \leq z \leq 2.5$ the shape of the observed LF is broadly reproduced 
by the SAM, the predicted characteristic density $\Phi^{\star}$ is
$\sim$2.5 times larger than the observed value. The SAM of De Lucia \& 
Blaizot has difficulty with both the normalization and the slope of the 
LF, which is too steep. The SPH simulations of \cite{finlator06} predict 
LFs that are qualitatively similar to those predicted by the two SAMs.

We quantified these results by determining the luminosity density
$j_{\rm V}$ (obtained by integrating the LF) for the observations and
models. The observed $j_{\rm V}$ ($j^{\rm obs}_{\rm V}$) has been 
estimated by integrating the best-fit Schechter function down to 
$M_{\rm V}=-19.5$, which is the faintest luminosity probed by the 
$K$-selected sample. To estimate $j_{\rm V}$ from the SAM 
($j^{\rm SAM}_{\rm V}$), we have fitted the predicted LFs with a 
Schechter function, leaving $M^{\star}$, $\Phi^{\star}$, and $\alpha$ 
as free parameters, applying the same limits as to the data.

\begin{figure}
\centering
\includegraphics[height=7cm]{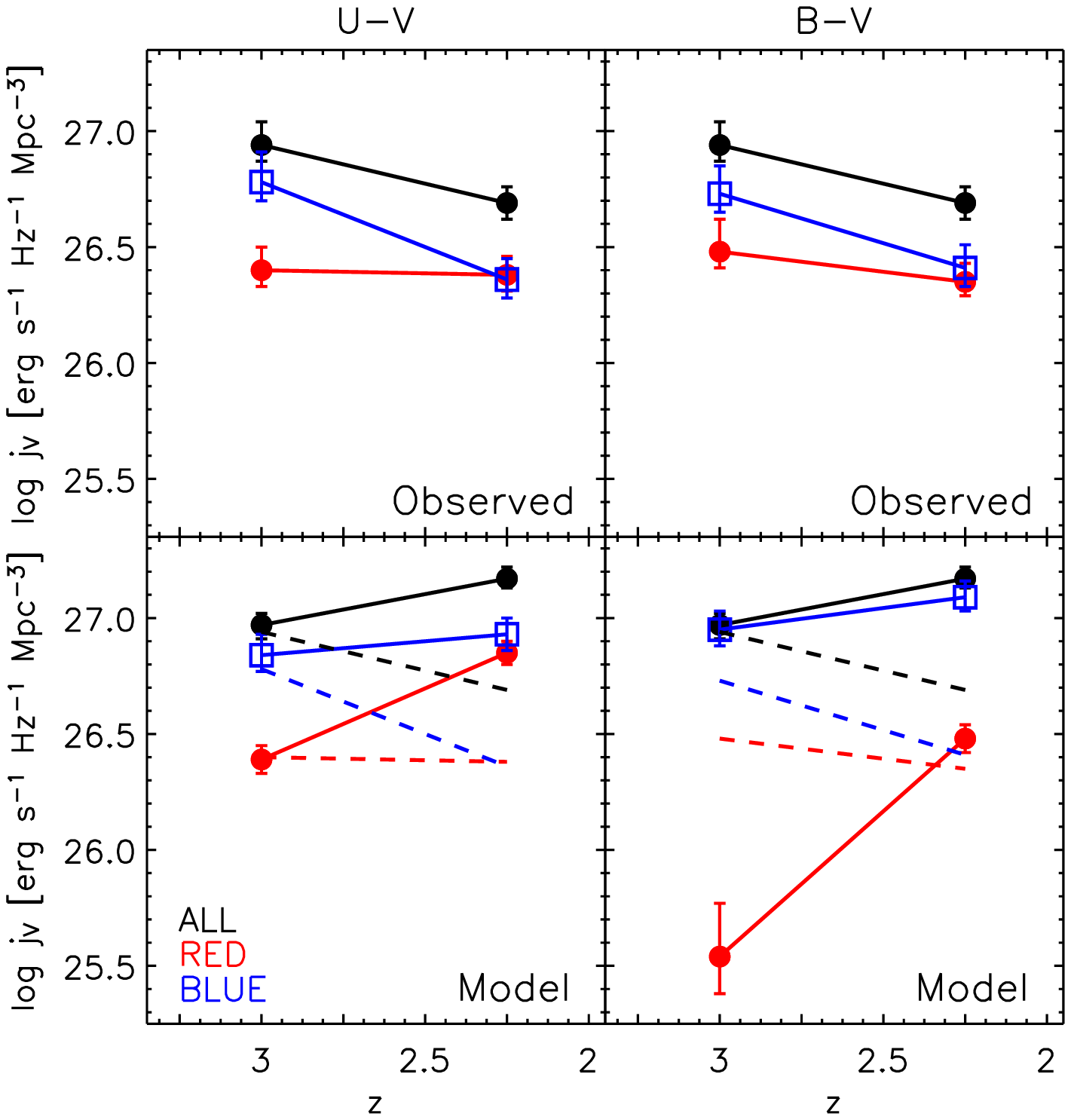}
\caption{From \cite{marchesini07b}: {\em Top panels:} Observed luminosity 
density ($j^{\rm obs}_{\rm V}$) as function of redshift of all 
({\it black circles}), red ({\it red circles}), and blue ({\it blue squares}) 
galaxies, splitting the sample based on rest-frame $U-V$ ({\it left panels}) 
and $B-V$ ({\it right panels}) colors. {\em Bottom panels:} Luminosity 
density predicted by the SAM of Bower et al. ($j^{\rm SAM}_{\rm V}$) 
as function of redshift; symbols as in top panels; the observed evolution
of $j_{\rm V}$ is also plotted with dashed lines for comparison.
\label{fig2}}
\end{figure}

The comparison between $j^{\rm obs}_{\rm V}$ and $j^{\rm SAM}_{\rm V}$
of Bower is shown in Figure~\ref{fig2} ({\it bottom panels}) 
by the black lines and data points. The Bower SAM matches the observed
luminosity density at $z\sim 3$. However, the model does not match the
evolution of $j_{\rm V}$. In the model the luminosity increases with
cosmic time, by a factor of $\sim1.6$ from $z\sim3$ to $2.2$. By
contrast, the observed luminosity density {\em decreases} with time,
by a factor of $\sim1.8$ over the same redshift range. Results for the
De Lucia SAM are similar, but for this model the difference between
observed and predicted density is a strong function of the adopted
faint-end integration limit.

\section{Colors}

As described in \cite{marchesini07b}, we investigated the cause of the 
discrepancies by splitting the sample into blue and red galaxies, using 
their rest-frame colors. Interestingly, the results depend strongly on 
the choice of color: splitting the sample by $U-V$ color (as done in
\cite{marchesini07}) produces very different results than splitting
by $B-V$ color.

To define red galaxies, we first use the criterion $U-V \geq$0.25, 
as done in \cite{marchesini07}.  As shown in the bottom left panel 
of Figure~\ref{fig2}, the Bower model reproduces the densities of 
red and blue galaxies at $z\sim 3$ extremely well, but it overpredicts 
the densities of red and blue galaxies at $z\sim 2.2$.

Next, we use the criterion $B-V \geq$0.5\footnote{For observed
galaxies in the sample of \cite{marchesini07}, $U-V=0.25$ roughly
corresponds to $B-V=0.5$.}. As can be seen in the top panels of
Figure~\ref{fig2}, this criterion leads to very similar observed
densities of red and blue galaxies as the $U-V$ criterion. However,
the predicted densities are in severe disagreement with the
observations, particularly at $z\sim 3$ (see Fig.~\ref{fig2}, 
{\it bottom right panel}). The red galaxy density at $z\sim 3$ underpredicts
the observed density by a factor of $\sim8$.  Qualitatively similar
results are obtained when $j^{\rm SAM}_{\rm V}$ from the SAM of
De~Lucia \& Blaizot is used in the comparison\footnote{The De Lucia
model provides $B-V$ colors, but no $U-V$ colors.}.

Irrespective of the color criterion that is used, we find that the
predicted {\em evolution} of the red and blue luminosity densities is
in qualitative disagreement with the observed evolution.  In the
observations, the moderate evolution of the luminosity density is
mainly driven by a decrease with cosmic time of the density of blue
galaxies, with the red galaxies evolving much less.  By contrast, in 
the SAMs, the moderate evolution seen in the global LF is in the 
opposite sense and dominated by a strong evolution in the red galaxy 
population.


\section{Discussion} \label{disc}

The main results of our comparison between the observed and the
model-predicted rest-frame $V$-band LFs of galaxies at $z \geq 2$ are 
(1) the SAM of Bower reproduces well the observed LF at 
$z\sim 3$; (2) the models predict an increase with time of the
rest-frame $V$-band luminosity density, whereas the observations show
a decrease; (3) the models predict strong evolution in the red galaxy
population, whereas in the observations most of the evolution is in
the blue population; (4) the models greatly underpredict the abundance
of galaxies with $B-V \geq0.5$ at $z\sim 3$.

The different results obtained for $U-V$ and $B-V$ colors are
interesting, as they may hint at possible ways to improve the models.
We further investigate the disagreement between observed and predicted
colors in the SAM of Bower by comparing the observed and predicted 
colors of galaxies in the $B-V$ versus $U-B$ diagram. While the SAM seems to 
broadly reproduce the observed $U-B$ distribution, it predicts galaxies 
that are systematically bluer in $B-V$ than the observed galaxies. 
As shown in \cite{marchesini07b}, the differences between observed and 
predicted colors could be due to larger amount of dust and/or to more 
complex star-formation histories in the observed galaxies. The ad hoc
treatment of dust absorption is a significant and well-known source of
uncertainty in the models.  Modifications to the specific dust model
could partly resolve the differences between observations and SAM
predictions. By simply multiplying the $A_{\rm V}$ in the SAM by a
fixed factor, we were able to better reproduce the observed LFs at
$z\sim2.2$ (although making the faint-end slope of the red galaxy LFs
quantitatively too steep) and to have a better agreement between
observed and predicted colors. However, at $z\sim3$ this simple remedy
is not able to solve the disagreement between the predicted and the 
observed number of $B-V \geq 0.5$ galaxies. We conclude that, while ad 
hoc modifications of the dust treatment might help to alleviate some 
of the found disagreements, it does not seem to be sufficient to 
resolve the problem with global colors at $z\sim3$.


\acknowledgments
D.M. is supported by NASA LTSA NNG04GE12G. The authors acknowledge 
support from NSF CARRER AST~04-49678.


\end{document}